\documentclass[12pt, a4paper]{article}
\usepackage[latin1]{inputenc}
\usepackage[dvips]{graphicx,color}
\usepackage{amsmath}
\usepackage{amsfonts}
\usepackage{amssymb}
\numberwithin{equation}{section}
\begin{document}
% Put preprint number in top-right.
\def\pplogo{\vbox{\kern-\headheight\kern -29pt
\halign{##&##\hfil\cr&{\ppnumber}\cr\rule{0pt}{2.5ex}&\ppdate\cr}}}
\makeatletter
\def\ps@firstpage{\ps@empty
\def\@oddhead{\hss\pplogo}
  \let\@evenhead\@oddhead}
\thispagestyle{firstpage} 
\def\maketitle{\par
 \begingroup
 \def\thefootnote{\fnsymbol{footnote}}
 \def\@makefnmark{\hbox{$^{\@thefnmark}$\hss}}
 \if@twocolumn
 \twocolumn[\@maketitle]
 \else \newpage
 \global\@topnum\z@ \@maketitle
\fi\thispagestyle{firstpage}\@thanks
 \endgroup
 \setcounter{footnote}{0}
 \let\maketitle\relax
 \let\@maketitle\relax

\gdef\@thanks{}\gdef\@author{}\gdef\@title{}\let\thanks\relax}
\makeatother

\newcommand{\nc}{\newcommand}
\def\theequation{\thesection.\arabic{equation}}
\nc{\beq}{\begin{equation}} \nc{\eeq}{\end{equation}}
\nc{\barray}{\begin{eqnarray}}
\nc{\earray}{\end{eqnarray}}
\nc{\barrayn}{\begin{eqnarray*}}
\nc{\earrayn}{\end{eqnarray*}}
\nc{\bcenter}{\begin{center}}
\nc{\ecenter}{\end{center}}
\nc{\ket}[1]{| #1 \rangle} \nc{\bra}[1]{\langle #1 |}
\nc{\0}{\ket{0}} \nc{\mc}{\mathcal} \nc{\etal}{{\em et
al}}
\nc{\GeV}{\mbox{GeV}} \nc{\er}[1]{(\ref{eq:#1})}
\nc{\onehalf}{\frac{1}{2}}
\nc{\partialbar}{\bar{\partial}}
\nc{\psit}{\widetilde{\psi}} \nc{\Tr}{\mbox{Tr}}
\nc{\vepsilon}{\varepsilon}

%%%%%%%%%%%%%%%%%%%%%%%%%%%%%%%%%%%%%%%%%%%%%%%%

\setcounter{page}0
\def\ppnumber{\vbox{\baselineskip14pt
%\hbox{hep-th/0000000}
}}
\def\ppdate{RUNHETC-2007-15} \date{}

\author{Kuver Sinha\\
[7mm]
{\normalsize NHETC, Department of Physics and Astronomy, Rutgers
University}\\
{\normalsize Piscataway, NJ 08855, U.S.A.}\\}

\title{\bf \LARGE Stabilizing the Runaway Quiver in
Supergravity}
\maketitle \vskip 1cm

\begin{abstract} \normalsize
\noindent We study stabilizations of the supersymmetry breaking runaway 
quiver in string embeddings. Calculations are performed in four dimensional effective supergravity. 
Constraints on closed string fields in a type IIA construction are given.
The particular case of stabilization by stringy instanton effects in a type IIB model is considered. 
\end{abstract}

\bigskip
\newpage

\tableofcontents

\vskip 1cm

%%%%%%%%%%%%%%%%%%%%%%%%%%%%%%%%%%%%%%%%%%%%%%%%%%%%%%%%%%%%%%%%%%%%%
%%%%%%%%%%%%%%%%%%%%%%%%%%%%%%%%%%%%%%%%%%%%%%%%%%%%%%%%%%%%%%%%%%%%%
\section{Introduction}\label{sec:intro}

It has been recently observed that fractional branes at singularities 
can give rise to quiver gauge theories that dynamically 
break supersymmetry. In \cite{beren}, the construction consisted of a 
cone over $dP_{1}$ and $N$ D5 branes at the end of the duality cascade 
with gauge group $U(3N) \times U(2N) \times U(N) $. It was argued that 
at the end of the cascade, confinement leads to deformation of the complex 
structure of the geometry; obstruction to such deformation causes 
supersymmetry breaking. A similar effect was conjectured for $Y^{p,q}, 
p > q >0$ and higher del Pezzos with the initial number of D5 branes 
chosen carefully. In \cite{franco1}, analysis was done in the regime 
where $U(3N)$ dominates, with the conclusion that the dynamically 
generated ADS superpotential drives the system away from the 
supersymmetric point at the origin of moduli space. A similar analysis was 
done for the case of $Y^{p,p-1}$ in \cite{bertolini} (see also 
\cite{franco2}, \cite{franco3}, \cite{Essig:2007xk},
\cite{franco4}, \cite{franco5}, \cite{franco6},
\cite{franco7}, \cite{Karp:2005ze}).

Supersymmetry breaking by this method has been used to engineer gauge mediation in string theory \cite{Diaconescu:2005pc}.
The standard model is realized using fractional branes on a partially collapsed $dP_{8}$, the supersymmetry breaking sector consists of branes at a
collapsed $dP_{1}$, and open strings stretching between the two stacks act as messangers.

However, in \cite{runaway}, Intriligator and Seiberg show by a detailed field 
theory analysis that the models proposed above have a runaway direction 
in field space, and thus do not actually break supersymmetry in the desired manner. In \cite{0603245}, dimer technology
was used to study the infrared behavior of the entire $Y^{p,q}$ family
and some examples of $L^{a,b,c}$ singularities, and arguments in favor of such runaway behavior were given.

Various constructions using open string fields have been made to stop 
this runaway and produce string-phenomenological models \cite{franco7}, \cite{GarciaEtxebarria:2006rw},
\cite{Argurio:2006ew}, \cite{Buican:2006sn}, \cite{GarciaEtxebarria:2007vh}. In \cite{franco7}, for example, extra fields 
were added to the model in the form of light massive flavors. By choosing a specific set of D7 branes, D3-D7 states were made to
couple with the fields of the D3 brane at the $dP_{1}$ singularity. The resulting extended quiver produced a long-lived metastable supersymmetry
breaking vacuum. Higher del Pezzos have been treated in \cite{Buican:2006sn}, while applications to gauge mediation have been
discussed in \cite{GarciaEtxebarria:2006rw}. Various other applications and extensions have been discussed in \cite{GarciaEtxebarria:2006aq},
\cite{Berenstein:2006vb}, \cite{Kachru:2007xp}, \cite{Wijnholt:2007vn}, \cite{Antebi:2007xw}, \cite{Malyshev:2007yb},
\cite{Kennaway:2007tq}, \cite{Argurio:2007vq}, \cite{Frampton:2007fr}, \cite{Kumar:2007dw}.

In this paper we consider supergravity stabilizations of the runaway quiver coming from closed string effects. 
 Closed string effects come from Calabi Yau moduli in realistic embeddings of the supersymmetry breaking quiver in string theory. 
Such embeddings have been done in both type IIA \cite{Diaconescu:2007ah} and type IIB \cite{Florea:2006si}.

At the level of the string embedding of the quiver, the runaway behavior in field space comes from the lack of proper moduli stabilization 
mechanisms. In the type IIA case, moduli stabilization is performed by RR and NS flux \cite{DeWolfe:2005uu}, \cite{Aharony:2007pr},
\cite{Jockers:2007ng}, \cite{Akerblom:2006hx}, \cite{Cvetic:2007ku}, \cite{Argurio:2007vqa}, \cite{Ibanez:2007rs}, \cite{Aharony:2007db}, 
\cite{Blumenhagen:2007bn}. Consistent orientifolding and the Freed Witten anomaly cancellation condition introduce various constraints 
on the Calabi Yau and the quiver locus. In this paper, we couple stabilized closed string moduli with the open string sector, and perform 
an effective four dimensional supergravity analysis for a variety of toy models. The conclusion is that under mild conditions on the 
Kahler potential and with proper choices of flux or instanton contributions to the superpotential, the quiver gauge theory is indeed 
stabilized. Comments on the possible uplift to $dS$ vacua are made. One expects these basic features to be true in a full-blown IIA computation.

In the case of embedding in type IIB, Kahler moduli stabilization comes from instanton effects \cite{Kachru:2003aw}. Embedding of the runaway 
quiver has been performed in \cite{Florea:2006si}, where various instanton effects have been explicitly calculated. We study this string realization in 
detail, and perform a supergravity analysis to show that the system stabilizes in a certain regime of calculability.

The plan of the paper is as follows. In section 2, we summarize the quiver gauge theory results in field theory. In section 3, we work out
general supergravity stabilization conditions, and apply them in a type IIA scenario. In section 4, we discuss stabilization in a type IIB construction, treating 
the example of \cite{Florea:2006si}. 

\textit{Acknowledgements. } I would like to thank my
advisor Emanuel Diaconescu for suggesting this 
problem and for discussions throughout. I would also
like to thank Gonzalo Torroba for discussions.

%%%%%%%%%%%%%%%%%%%%%%%%%%%%%%%%%%%%%%%%%%%%%%%%%%%%%%%%%%%%%%%%%%%%
%%%%%%%%%%%%%%%%%%%%%%%%%%%%%%%%%%%%%%%%%%%%%%%%%%%%%%%%%%%%%%%%%%%%

\section{The Runaway Quiver : Field Theory
 Description}\label{sec:field}

The gauge theory of $M$ D5 branes on the complex cone
over $F_1$ is given by
$SU(3M) \times SU(2M) \times SU(M)$. For the purpose
of this paper, we consider the case $M = 1$.
The various fields transform as follows \cite{runaway}:

\begin{equation}\label{eq:gaugetheory} 
\begin{matrix} 
& SU(3M) & \, \, \, \, \, \, SU(2M) & \, \, \, \, \, \, SU(M) & \, \, \, \, \, \, [ SU(2) & \, \, \, \, \, \, U(1)_F & \, \, \, \, \, \, U(1)_R
] \cr \cr
 Q & {\bf 3M} & {\bf{\overline{2M}}} & {\bf 1} & \, \, \, {\bf
1} & \, \, \, 1 & \, \, \, -1\cr\cr
 \overline u & {\bf{\overline{3M}}} & {\bf 1} & {\bf
M} & \, \, \, {\bf 2} & \, \, \, -1
 & \, \, \, 0\cr\cr
 L & {\bf 1} & {\bf 2M} & {\bf{\overline{M}}} & \, \, \, {\bf
2} & \, \, \, 0 & \, \, \, 3 \cr\cr
 L_3 & {\bf 1} & {\bf 2M} & {\bf{\overline{M}}} & \, \, \, {\bf
1} & \, \, \, -3 & \, \, \, -1
\end{matrix}
\end{equation}

The gauge invariant fields are defined as

\begin{equation}\label{eq:gaugeinvts}
Z=\det_{fj} Q^f\overline u_j,
 \quad X_{ia}=Q\overline u_i L_a,
 \quad V^a={1\over 2}L_bL_c\epsilon ^{abc}.
\end{equation}

The low energy spectrum of the system consists of the fields $V^i, \, i = 1,2$ and $V^{3}
 \equiv V$  after all other fields have satisfied their SUSY equations of motion.
 The dynamical superpotential is

\begin{equation}\label{eq:Wfull}
W \,= \, 3 \, (V \, \Lambda _3^7)^{1/3}.
\end{equation}
The Kahler potential far out in $V$ moduli space is given by

\begin{equation}\label{eq: Keff}
K_{eff}\approx K_{cl}=2\sqrt{T}=2 \sqrt{ VV^\dagger
 +V^iV^i{}^\dagger }
\end{equation}
This leads to a runaway in field space of the form:

\begin{equation}\label{eq:runlolarun}
V_{eff}\approx 2|
 \Lambda_3^7|^{2/3}(VV^\dagger)^{-1/6}
\end{equation}
with $V_i = 0 $.

%%%%%%%%%%%%%%%%%%%%%%%%%%%%%%%%%%%%%%%%%%%%%%%%%%%%%%%%%%%%%%%%%%%%%%%%%%%%%%%%%

%%%%%%%%%%%%%%%%%%%%%%%%%%%%%%%%%%%%%%%%%%%%%%%%%%%%%%%%%%%%%%%%%%%%%%%%%%%%%%%%%%

\section{Stabilization Conditions and Type IIA examples }\label{sec:decoup}

In embeddings of the above quiver gauge theory in type IIB string theory, the runaway in field space is caused by a lack of moduli stabilization 
mechanism at the string level. Closed string Kahler moduli are typically stabilized by instanton effects. However, in the present context, such instantons develop extra
zero modes due to their interaction with the fractional branes, which can lead to cancellations in the effective superpotential. Some progress has been made 
recently in that direction, for example in \cite{Florea:2006si}, where instanton effects are explicitly calculated. In the next section, we consider stabilizations in such a 
scenario.

An alternate embedding begins with the observation that such quiver gauge theories occur at non-geometric phases in the 
Kahler moduli space, and hence can be treated in a type IIA scenario by using mirror symmetry. Supergravity methods can be used in the mirror picture. 
A full blown IIA embedding of the runaway quiver 
consists of generic NS and RR flux stabilizing complex structure and Kahler moduli of the Calabi Yau $Y$ respectively. The quiver is 
realized by D6 branes wrapping special Lagrangian cycles. In compact models, an orientifold projection is introduced. A number of conditions on 
the quiver locus and the geometry of the IIB mirror Calabi Yau $X$ have to be imposed -  $(i)$ $X$ should contain a pair of disjoint del Pezzos $(S,S')$ which don't 
intersect the fixed point locus $X^{\sigma}$ of the orientifold projection and $(ii)$ the holomorphic involution of the orientifold projection should be 
compatible with the large complex structure limit in the complex structure moduli space of $X$, so that computations can be done in the supergravity limit 
of the mirror IIA scenario.

In the mirror IIA construction, the superpotential gets the following flux contributions

\begin{equation} \label{eq: IIAfluxes}
W_{K} \, = \, \int_{Y} F \wedge e^{-J_Y}, \, \, \, \,  W_H(x^k, t_{\lambda}) \, = \, -2x^kg_k - it_{\lambda}h^{\lambda}
\end{equation}
where $W_K$ is the superpotential contribution to the Khaler moduli of $Y$, $J_Y$ is the Kahler class of $Y$, $F$ is the RR flux and is given by 
$F = F_0 + F_2 + F_4 + F_6$, $W_H$ is the superpotential contribution to the complex structure moduli of $Y$, $(g_k,h^{\lambda})$ are the NS flux, and 
$x^k,t_{\lambda}$ are $h^3_{+} = h^{2,1} + 1 $ holomorphic coordinates on the $N=1$ complex structure moduli space \cite{Grimm:2004ua}, \cite{Diaconescu:2006nk}. 
The coordinates $x^k$ and $t_\lambda$ are given by 

\begin{equation} \label{eq: holocoords}
x^{i} \, = \, 1/2 \int_Y \Omega_Y^c \wedge \beta^i, \, \, \, \,  t_\lambda \, = \,  \int_Y \Omega_Y^c \wedge \alpha_{\lambda}
\end{equation}
where $(\alpha_\lambda,\beta^i)$ form a symplectic basis of three-cycles on $Y$ and $\Omega_Y^c$ is a linear combination of the RR three-form $C^{(3)}$ and the real part 
of the holomorphic three-form of $Y$. A specific choice of symplectic basis for explicit calculations demands more constraints on the construction \cite{Diaconescu:2007ah} 
- $(iii)$ the natural push-forward maps $H_2(S) \rightarrow H_2(X)$ and $H_2(S') \rightarrow H_2(X)$ 
have rank one, and $(iv)$ under the orientifold projection, the anti-invariant subspace $H_{-}^{1,1}(X)$ is one-dimensional and is spanned by the difference $S-S'$ 
between the divisor classes of the conjugate del Pezzos $S$ and $S'$.

In general, $W_K$ is enough to stabilize all Kahler moduli. On the other hand, NS flux is subject to the Freed-Witten anomaly cancellation condition, 
which can hinder moduli stabilization by hindering generic flux. Requiring F-flatness of the superpotential then requires additional conditions 
due to the non-appearance of certain complex structure moduli due to the anomaly cancellation condition. In \cite{Diaconescu:2007ah}, explicit embeddings of the 
quiver gauge theory have been constructed taking into account all the above constraints in the case of certain quintic threefolds. 

In this section, we perform an effective four-dimensional supergravity analysis of the quiver. We take general closed string contributions to the Kahler potential and 
superpotential, and couple them to the open string sector. Our strategy is to begin with a supersymmetric vacuum on the closed string side, and stabilize the 
open string field $\psi \, = \, \kappa^2 V_3$ in that vacuum. A self-consistent analysis is performed for $\psi \ll 1$, which allows independent stabilization of the closed string 
sector and removes higher order corrections to the open string sector coming from $U(1)$ D-terms. Comments on the possible uplift to $dS$ vacua are made. 

We work out specific examples for the case of a single complex modulus $x$ in a type IIA context, without taking into account the complications introduced by the Freed Witten anomaly. The Kahler 
potential is taken to be a power series in $x$, while the superpotential is considered to be either a flux contribution like (\ref {eq: IIAfluxes}), or a typical instanton effect. 

Our general result is that in the case of a flux superpotential, tuning the value of flux enables stabilization in the region of calculability and possible uplift to small 
positive cosmological constant. In the case of an instanton superpotential, consistent stabilization without strong constraints on the Kahler potential or superpotential 
requires a hierarchy of scales between the two sectors. Uplift to $dS$ vacuum is correspondingly more difficult to achieve. The stabilization procedure in both cases puts mild 
conditions on the Kahler potential, and in the second case, on the instanton contribution. It is expected that in a full IIA calculation, these basic features would be maintained.

%%%%%%%%%%%%%%%%%%%%%%%%%%%%%%%%%%%%%%%%%%%%%%%%%%%%%%%%%%%%%%%%%%%%%%%%%%%%%%%%%%%%%%%%%%%%%%%%%%%%%%%%%%%%%%%%%%%%%%%%%%%%%%%%%%
%%%%%%%%%%%%%%%%%%%%%%%%%%%%%%%%%%%%%%%%%%%%%%%%%%%%%%%%%%%%%%%%%%%%%%%%%%%%%%%%%%%%%%%%%%%%%%%%%%%%%%%%%%%%%%%%%%%%%%%%%%%%%%%%%%
%%%%%%%%%%%%%%%%%%%%%%%%%%%%%%%%%%%%%%%%%%%%%%%%%%%%%%%%%%%%%%%%%%%%%%%%%%%%%%%%%%%%%%%%%%%%%%%%%%%%%%%%%%%%%%%%%%%%%%%%%%%%%%%%%%

\subsection{General Analysis}\label{sec:general3}

We take the following Kahler potential and superpotential:

\begin{equation} \label{eq: K 2}
\kappa^2 K = (\psi \bar{\psi})^{1/2} \, (1 + \gamma
\sum p_i ) + \sum
 f_i
\end{equation}

\begin{equation} \label{eq:W 2}
W = \Lambda^3 \psi^{1/3} \, (1+ \sigma \sum q_i)  +
\Lambda_1 ^3 \sum
 g_i
\end{equation}
where $p_i = p(x_i, \bar x_i), \, f_i = f(x_i, \bar
x_i), \, q_i = q(x_i), \, g_i = g(x_i)$, the $x_i$ being an
arbitrary number of closed string moduli. We consider the simple case where different closed
string moduli $x_i$ and $x_j$ are decoupled. $\gamma$ and $\sigma$ parametrize the strength of the coupling between the open and closed string sectors in the
Kahler potential and superpotential respectively. In particular, we will be working to first order in these parameters. Also, $\psi = \kappa^2 V_3 \ll 1 $, so that
the field $V_3$ is stabilized below the Planck scale.

The Kahler metric may be inverted, and to first order in $\gamma$ one obtains

$$
K^{\psi \bar \psi} = 4 \kappa^2 \, (1 - \gamma \sum
p_i) \,  \vert \psi
 \vert
$$

$$
K^{x_i \bar \psi} = (1/2) \kappa^2 \, \gamma \, 
\partial_i p_i \,
 (\partial_i \bar \partial_i f_i)^{-1} \, \psi
$$

\begin{equation} \label{eq: K inverse}
K^{x_i \bar{x_i}} =  \kappa^2 \,\bigl[ (\partial_i
\bar \partial_i
 f_i)^{-1} \, - \, \gamma  \, (\partial_i \bar
\partial_i p_i) \,
  (\partial_i \bar \partial_i f_i)^{-2} \, \vert \psi
\vert \bigr]
\end{equation}
while $K^{x_i \bar x_j} , i \neq j$ starts at order
$\gamma^2$.

All the contributions to the supergravity scalar potential can be computed, and we keep terms upto order $\vert \psi
\vert ^{1/3}$. The resulting stabilization places constraints on the
functions $f_i, g_i, p_i, q_i$. Generally, the potential is of the form

\begin{equation} \label {eq:V 2}
V = e^{\kappa^2 \, \Sigma f} \biggl[ A \vert \psi \vert ^{1/3} \, \, + B \vert \psi \vert
^{-1/3} \biggr] +
 V_{\rm{closed}}
\end{equation}
where $A,B$ can be expressed in terms of 
$f_i, g_i, p_i,
 q_i$. 

The non-Abelian D-term contributions to the potential are set to 
zero by working on the D-flat moduli space defined by (\ref{eq:gaugeinvts}).
The $U(1)$ D-term contributions in general introduce new open-closed mixing 
terms, since the gauge coupling is a holomorphic function of the closed string 
moduli. However, these mixings begin at order $\vert \psi \vert$, and we neglect them.

We study some limiting cases of the parameters $\gamma$ and $\sigma$, and work out some examples 
with a single complex structure modulus. The functions $f$ and $p$ in the Kahler 
potential are taken to be power series expansions in the complex structure modulus. The superpotential
term is taken to be a flux contribution or a typical instanton contribution.

We note that similar supergravity calculations have been performed (see \cite{Kallosh:2006dv}, 
\cite{Abe:2007yb}, for example) in the context of uplifting the KKLT $AdS$ vacuum by coupling it to a
SUSY breaking sector such as an O'Raifeartaigh or ISS model.

%%%%%%%%%%%%%%%%%%%%%%%%%%%%%%%%%%%%%%%%%%%%%%%%%%%%%%%%%%%%%%%%%%%%%%%%%%%%%%%%%%%%%%%%%%%%%%%%%%   
%%%%%%%%%%%%%%%%%%%%%%%%%%%%%%%%%%%%%%%%%%%%%%%%%%%%%%%%%%%%%%%%%%%%%%%%%%%%%%%%%%%%%%%%%%%%%%%%%%
%%%%%%%%%%%%%%%%%%%%%%%%%%%%%%%%%%%%%%%%%%%%%%%%%%%%%%%%%%%%%%%%%%%%%%%%%%%%%%%%%%%%%%%%%%%%%%%%%

\subsection{$\gamma = \sigma = 0$}\label{subsec:gammasigma0}

In this case, one obtains 

$$
A = \sum_i \, \, \kappa^2 \Lambda^3 \Lambda_1 ^3 \,
\biggr[ (2/3)\Sigma
 \, \bar g \, + \, [\partial_i f_i \bar \partial_i
\bar {g_i} +
 \partial_i f_i \bar \partial_i f_i \Sigma \, \bar g]/
\partial_i \bar
 \partial_i f_i \,   \, - 3 \Sigma \, \bar g \biggl] 
e^{i\theta/3} + c.c.
$$

\begin{equation} \label {eq: Bsubsec3.1}
B = (4/9) \kappa ^2 \Lambda ^6  
\end{equation}
Here, $\theta$ is the phase of $\psi$. We work out the case of a single
complex structure  modulus $x$, with  $ f = f_0 \, + \, \alpha_1 (x +
 \bar x) \, + \, \alpha_2(x \bar x) + ...$.

$(i)$ Taking a single complex structure modulus, we have $g = g_0x$. For $\vert \psi \vert \ll 1$ we can stabilize the closed string sector independently. A stable 
supersymmetric solution is located at 

$$
{\rm Re} x_{{\rm min}} \, = \, ({\rm Re}\alpha_1/2\alpha_2)\bigr[ -1 \pm \sqrt{1 - (4\alpha_2/\alpha_1^2)} \bigl], 
$$

\begin{equation} \label{eq: xminflux3.1}
{\rm Im} x_{{\rm min}} \, = \, -({\rm Im}\alpha_1/{\rm Re}\alpha_1) {\rm Re} x_{{\rm min}}.
\end{equation}
Stabilizing the open string sector, one obtains 

\begin{equation} \label{eq : psistabflux3.1}
\vert \psi \vert_0^{1/3} \, = \, [(4/9)(\Lambda/\Lambda_1)^3]^{1/2} \, g_0^{-1/2} \, J^{-1/2}
\end{equation}
where

\begin{equation} \label{eq : condcoeff3.1}
J \, = \, (4/3 \,  - \, \alpha_1^2/\alpha_2)({\rm Re}x_{{\rm min}}) \, -  \, {\rm Re}\alpha_1/\alpha_2 \, - \, 4\alpha_1 x_{\rm min}^2 \, g_0 > 0.
\end{equation}
Note that (\ref{eq : condcoeff3.1}) places constraints on the coefficients appearing in the Kahler potential. Minimization with respect to phases has been done, 
and for simplicity we have assumed $\text{Im} \alpha_1 $ is small.
The self-consistency condition $\vert \psi \vert \ll 1$ can be obtained by tuning the flux $g_0$ to be large, without assuming a hierarchy of scales between $\Lambda$ and $\Lambda_1$.
On the other hand, the value of the potential at the minimum is 

\begin{equation} \label{eq : Vminflux3.1}
V_{\rm{min}} \, = \, 2e^{\kappa^2 K}\kappa^2 \Lambda^{9/2}\Lambda_1^{3/2}g_0^{1/2}J^{1/2} \, - \, 3 \kappa^2 e^{\kappa^2K}g_0^2 x_{\rm{min}}^2 \Lambda_1^6.
\end{equation}
Tuning the flux such that $J \, \sim \, g_0 x_{\rm{min}}^2$, one can potentially lift to a $dS$ vacuum.

On the other hand, assuming a hierarchy of scales $\Lambda/\Lambda_1 \ll 1$ without tuning the flux automatically satisfies $\vert \psi \vert \ll 1$, 
but in this case uplift to a $dS$ vacuum is difficult to achieve.

$(ii)$ Taking a typical instanton correction to the superpotential sets $g = \beta e^{-\alpha x}$. For 
$\vert \psi \vert \ll 1$, the stabilization of the closed string sector is decoupled from the open string sector.
We start with a stable closed sting vacuum satisfying $D_x W = 0$, located at

\begin{equation} \label {eq : xmininst3.1}
x_{\rm {min}} \, = \, (1/\alpha_2)(\alpha - \bar {\alpha_1})
\end{equation}
For small
$ \text{Im} x$, minimizing with respect 
to the phases sets $(\theta/3 \, + \, \alpha
\text{Im}\ x) \,= \pi$. 
Minimizing the open string sector with respect to $\vert \psi \vert$ sets 

\begin{equation} \label{eq: psimininst3.1}
\vert \psi \vert _0 ^{1/3} \, \sim \, (4/21)^{1/2}
(\Lambda/\Lambda_1)^{3/2}
 \, \beta^{-1/2} \, \rm{exp}[(\alpha/2\alpha_2)(\alpha - \alpha_1/2 - \bar \alpha_1/2)]
\end{equation}
The condition $\vert \psi
\vert \ll 1$ can be achieved by having $\beta \gg 1$ and $\Lambda \ll \Lambda_1$.

At the minimum, we obtain 

$$
V \, \sim \, (112/27)^{1/2} \kappa^2 \, \Lambda^3 \, (\Lambda
\Lambda_1)^{3/2}
 \, \beta^{1/2} \, \rm{exp} [f_0 - (\alpha - \alpha_1)^2/\alpha_2] \times
$$
\begin{equation} \label{eq : Vmininst3.1}
\times \, \rm{exp}[(3\alpha/2\alpha_2)(\alpha - \alpha_1/2 - \bar \alpha_1/2] \, - \,
3 \kappa^2 \Lambda_1^6 \beta^2 \rm{exp}[f_0 - (\alpha - \alpha_1)^2/\alpha_2]
\end{equation}
In the regime of calculability $\vert \psi \vert \ll 1$, the vacuum remains close to the
closed string $AdS$ vacuum, and there isn't much uplift.

%%%%%%%%%%%%%%%%%%%%%%%%%%%%%%%%%%%%%%%%%%%%%%%%%%%%%%%%%%%%%%%%%%%%%%%%%%%%%%%%%%%%%%%%%%%%%%%%%%%%%%%%%%%%%%%%%%%%%%%%%%%%%%%%%%%%%%%%%%%%%%%%%%%%%%%%%%%%%%%%%%%%%
%%%%%%%%%%%%%%%%%%%%%%%%%%%%%%%%%%%%%%%%%%%%%%%%%%%%%%%%%%%%%%%%%%%%%%%%%%%%%%%%%%%%%%%%%%%%%%%%%%%%%%%%%%%%%%%%%%%%%%%%%%%%%%%%%%%%%%%%%%%%%%%%%%%%%%%%%%%%%%%%%%%%%
%%%%%%%%%%%%%%%%%%%%%%%%%%%%%%%%%%%%%%%%%%%%%%%%%%%%%%%%%%%%%%%%%%%%%%%%%%%%%%%%%%%%%%%%%%%%%%%%%%%%%%%%%%%%%%%%%%%%%%%%%%%%%%%%%%%%%%%%%%%%%%%%%%%%%%%%%%%%%%%%%%%%%

\subsection{$\gamma = 0, \sigma \neq 0$} \label{subsec:g0sign0}

In general, apart from flux contributions to the superpotential, instanton corrections coming from the closed string sector can couple to the open string fields. In that 
case, $\sigma \neq 0, \, q \, = \, \beta e^{-\alpha x} $. Such corrections will also lead to open-closed coupling in the Kahler potential, but as a limiting case we set $\gamma = 0 $ here. We take the case 
of a single IIA complex structure modulus and consider two cases - where the pure closed string contribution $g$ is a flux effect, and where $g$ is also due to an instanton 
effect. 

In section \ref{sec:inst}, we study a type IIB embedding scenario where such couplings have been explicitly calculated. 

For $\gamma = 0, \sigma \neq 0$, we obtain

$$
A = \sum_i \, \, \kappa^2 \Lambda^3 \Lambda_1 ^3 \,
\biggr[ (\sigma
 \bar \partial_i \bar g_i \partial_i q_i)/\partial_i
\bar \partial_i  f_i
 \, + \, (2/3)\Sigma \bar g \, 
+ \, [\partial_i f_i \bar \partial_i \bar g_i (1 +
\sigma \Sigma q) \,
 +  
$$
$$
+ \sigma \bar \partial_i f_i \partial_i q_i \Sigma
\bar g]/\partial_i
 \bar \partial_i f_i  \, + \, (1 + \sigma \Sigma
q)\partial_i f \bar
 \partial_i f \Sigma \bar g / \partial_i \bar
\partial_i  f_i \, - \,3 (1 +
 \sigma \Sigma q)\Sigma \bar g \biggl] e^{i\theta/3} +
c.c.
$$

\begin{equation} \label{eq:gamma0}
B = (4/9) \kappa^2 \Lambda^6 \, (1+ \sigma \Sigma q)
\end{equation}

$(i)$ We consider a flux contribution to the superpotential as before $g=g_0 x$ and take $q = \beta e^{-\alpha x}$. For $\vert \psi \vert \ll 1$ and  $\sigma$ such that 
$\vert \psi \vert ^{-1/3}\sigma \ll 1$, the 
open-closed mixing in the potential is small, and the closed string sector can be stabilized independently as before.
We obtain a  supersymmetric minimum, where the value of $x$ is given by (\ref{eq: xminflux3.1}). Stabilization on the 
open string side gives $\psi$ as a function of the coefficients $\alpha, \alpha_1, \alpha_2$. For small $x$, this simplifies and we get

\begin{equation} \label{eq:psimin3.2}
\vert \psi \vert ^{1/3} \, = \, (2/3)(\Lambda/\Lambda_1)^{3/2} \, g_0^{-1/2} \, [\alpha_2 ( 1 + \sigma \beta)]^{1/2}[\sigma\beta\alpha - (1 + \sigma\beta) \rm{Re}\alpha_1]^{-1/2}.
\end{equation}
The calculability condition can be satisfied by taking large values of $g_0$. We also note that reality of $\vert \psi \vert$ sets the condition 
$\sigma\beta\alpha - (1 + \sigma\beta) \rm{Re}\alpha_1 \, > \, 0$. The value of the potential at the minimum is given by 
$$
V_{\rm{min}} \, = \,  2e^{\kappa^2 K}\kappa^2 \Lambda^{9/2}\Lambda_1^{3/2} \, g_0^{1/2} \,  [\alpha_2 ( 1 + \sigma \beta)]^{1/2}[\sigma\beta\alpha - 
(1 + \sigma\beta) \rm{Re}\alpha_1]^{1/2} \, - 
$$
\begin{equation} \label{eq: vmin3.2}
- \, 3 \kappa^2 \, \Lambda_1^6 \, e^{\kappa^2K} \, g_0^2 \, x_{\rm{min}}^2 .
\end{equation}
In principle, it is possible to uplift the $AdS$ vacuum by controlling the flux $g_0$ such that $(g_0^{3/2}x_{\rm{min}}^2)^{-1} \sim 1$.

$(ii)$ We now consider the case where the pure closed string contribution to the superpotential is also an instanton effect. In this case, 
$g = \beta_g \, e^{-\alpha x}$,  $q = \beta_q \, e^{-\alpha x}$. The supersymmetric minimum of the closed string sector is given by (\ref{eq : xmininst3.1}). 
For small $x$, the open string sector is stabilized at 

\begin{equation} \label{eq: psimininst3.2}
\vert \psi \vert ^{1/3} \, = \, (2/3)(\Lambda/\Lambda_1)^{3/2} \, [1+\sigma \beta_q]^{1/2} \, J^{-1/2},
\end{equation}
where $J\, = \, \beta_g[7/3 + (\alpha\alpha_1/\alpha_2) - \alpha_1^2/\alpha_2] \, + \, \beta_g \beta_q 
[3\sigma + 2\sigma\alpha\alpha_1/\alpha_2 - \sigma\alpha^2/\alpha_2 - \sigma\alpha_1^2/\alpha_2] > 0$ is a condition that can be satisfied if $\alpha \, > \, \alpha_1$, 
for example. Also, $\vert \psi \vert \ll 1 $ requires the hierarchy of scales $\Lambda/\Lambda_1 \ll 1$.

As in (\ref{eq : Vmininst3.1}), the minimum of the system remains close to the $AdS$.

%%%%%%%%%%%%%%%%%%%%%%%%%%%%%%%%%%%%%%%%%%%%%%%%%%%%%%%%%%%%%%%%%%%%%%%%%%%%%%%%%%%%%%%%%%%%%%%%%%%%%%%%%%%%%%%%%%%%%%%%%%%%%%%%%%%%%%%%%%%%%
%%%%%%%%%%%%%%%%%%%%%%%%%%%%%%%%%%%%%%%%%%%%%%%%%%%%%%%%%%%%%%%%%%%%%%%%%%%%%%%%%%%%%%%%%%%%%%%%%%%%%%%%%%%%%%%%%%%%%%%%%%%%%%%%%%%%%%%%%%%%%
%%%%%%%%%%%%%%%%%%%%%%%%%%%%%%%%%%%%%%%%%%%%%%%%%%%%%%%%%%%%%%%%%%%%%%%%%%%%%%%%%%%%%%%%%%%%%%%%%%%%%%%%%%%%%%%%%%%%%%%%%%%%%%%%%%%%%%%%%%%%%

\subsection{$\gamma \neq 0, \sigma = 0$ } \label{subsec :gamn0sig0}

In the limit where open and closed string contributions may be taken to be decoupled in the superpotential, the Kahler potential of the system 
will in general still contains couplings between the two sectors. Considering $\gamma \neq 0, \sigma = 0 $, we get

$$
A = \sum_i \, \, \kappa^2 \Lambda^3 \Lambda_1 ^3 \,
\biggr[ (2/3)\Sigma
 \, \bar g \, + \, [\partial_i f_i \bar \partial_i
\bar {g_i} +
 \partial_i f_i \bar \partial_i f_i \Sigma \, \bar g]/
\partial_i
 \partial_{\bar i} f_i \,   \, - \, 3 \Sigma \, \bar g
\, - 
$$
$$
- \, (2/3) \, \gamma \, \partial_i \bar g_i \partial_i
p_i /\partial_i
 \bar \partial_i f_i
\biggl]  e^{i\theta/3} + c.c
$$

\begin{equation} \label{eq : sigma0}
B = (4/9) \kappa ^2 \Lambda ^6 (1-\gamma \Sigma p)  
\end{equation}

$(i)$ We take $g = g_0x$, $ f = f_0 \, + \, \alpha_{1f} (x + \bar x) \, + \, \alpha_{2f}(x \bar x) + ...$, 
and $ p = p_0 \, + \, \alpha_{1p} (x + \bar x) \, + \, \alpha_{2p}(x \bar x) + ...$. The closed string sector is 
stabilized at the supersymmetric vacuum given by (\ref{eq: xminflux3.1}). For $x \rightarrow 0$, the open string sector is 
stabilized at 

\begin{equation} \label{eq: psimingneq0sig0}
\vert \psi \vert ^{1/3} \, = \, (2/3)(\Lambda/\Lambda_1)^{3/2} \, g_0^{-1/2} \,[1-\gamma p_0]^{1/2} \, [2/3 \gamma \alpha_{1p}/\alpha_{2f} - \alpha_{1f}/\alpha_{2f}]^{-1/2}.
\end{equation}
This gives the constraint $2/3 \gamma \alpha_{1p} \, > \, \alpha_{1f}$. As before, $\vert \psi \vert \ll 1 $ can be achieved by $g_0 \gg 1$, while an uplift of the 
$AdS$ vacuum can be achieved by tuning $g_0$ such that $(g_0^{3/2}x_{\rm{min}}^2)^{-1} \sim 1$.

$(ii)$ For an instanton-like contribution $g = \beta e^{-\alpha x}$, the closed string supersymmetric minimum lies at (\ref{eq: xminflux3.1}), while for small $x$, 
the open string field is stabilized at 

\begin{equation} \label{eq: psiminlast}
\vert \psi \vert _0 ^{1/3} \, = \, (4/3) (\Lambda/\Lambda_1)^{3/2} \, \beta^{-1/2} \, (1-\gamma p_0)^{1/2} \, J^{-1/2}
\end{equation}
where $J \, = \, (7/3) - \, (\alpha_{1f}^2/\alpha_{2f}) \, +   \,  \alpha \alpha_{1f}/ \alpha_{2f} - \, (2/3) \gamma \alpha \alpha_{1p}/\alpha_{2f}$. We require 
$J > 0$.

%%%%%%%%%%%%%%%%%%%%%%%%%%%%%%%%%%%%%%%%%%%%%%%%%%%%%%%%%%%%%%%%%%%%%
%%%%%%%%%%%%%%%%%%%%%%%%%%%%%%%%%%%%%%%%%%%%%%%%%%%%%%%%%%%%%%%%%%%%%
\section{Stabilization with Stringy
Instantons in IIB}\label{sec:inst}

Following \cite{Florea:2006si}, we consider the quiver gauge theory
 on a singular $dP_{1}$ geometry, with added Euclidean D3 brane
 instantons. The D3's which intersect the singularity will in general also give
 rise to Ganor strings stretching from the occupied nodes of the
 quiver. Denoting quiver fields generically by $\psi_{i}$, the superpotential
 of the system is deformed by effects of the form 

\begin{equation} \label{eq:gendef}
\Delta W \sim f(\psi_{i})\exp (- \text{Vol})
\end{equation}
where Vol is the volume of the D3. The scalar
potential can be stabilized to obtain metastable vacua.

Concretely, the complex cone over $dP_{1}$ can be described in terms of
 toric data as follows. The non-trivial two-cycles in
$dP_{1}$ are denoted by $f$ and $C_{0}$. A basis of branes is given by
\begin{equation} \label{eq:basis}
[\, \mathcal{L}_1, \mathcal{L}_2, \mathcal{L}_3,
\mathcal{L}_4 \, ] =
\, [ \mathcal{O}_{{F}_1},
\mathcal{O}_{F_1}(C_0 + f),
\overline{\mathcal{O}_{F_1}(f)},
\overline{ \mathcal{O}_{F_1}(C_0)} \,].
\end{equation}
Denoting the $\mathbb{P}^{1}$ fibrations over $f$ and $C_{0}$
by $D_{2}$ and $D_{3}$, and the the $dP_{1}$ base by $D_{5}$, one obtains the nonzero
triple intersections

\begin{equation} \label{eq:intnos}
D_{5}^3 = 8 , \, \, D_{5}D_{2}D_{3} = 1, \, \, D_{5}^2
D_{2} = D_5
 D_{3}^2 = -1 \,. 
\end{equation}

Various instanton effects can be calculated in this geometry. This
 requires knowledge about the topology of the D3 brane
and its spectrum of Ganor strings. The Euclidean D3's and the quiver
nodes wrap a surface $S$ on the del Pezzo cone, and carry different line
bundles $\mathcal{L}_{A}$ and $\mathcal{L}_{B}$ over $S$. The
most general bundle for the instanton is $X_{ab} = {\overline{ {\cal
O}_{D_5}(aC_0 + bf)}}$. Computing the number of fermionic zero modes between $X_{ab}$
and $\mathcal{L}_{1,2,3}$ gives

\begin{equation} \label{ eq: zeroes}
n_{ \text{ferm} }(X_{ab},\mathcal{L}_{1,2,3}) =
(a+2b,~-3+a+2b,~2-a-2b) \,.
\end{equation}

An important instanton effect one can have in this
geometry is the Affleck-Dine-Seiberg (ADS) instanton effect. In this
case, $a=0, \, b=1$; that is, the instanton wraps the bundle
$\mathcal{L}_3$. It turns out that the instanton contribution in this case leads to
the superpotential

\begin{equation} \label{eq: Wads}
W_{ADS}={\Lambda^7 \over Z}e^{-S_1}.
\end{equation}
Here, $S_1$ is given by

\begin{equation} \label{eq:S_1}
{ \rm Re} (S_1)=
(1/2(8r_5^2-r_3^2-2r_3r_5-4r_2r_5+2r_2r_3)+r_3-2r_5.
\end{equation}
where $r_5, r_2, r_3$ parametrize the Kahler form $J$
in terms of the toric data in the following way:

\begin{equation} \label{eq:J}
J = r_{5}D_{5} \, + r_{2}D_{2} \, + r_{3}D_{3} \, .
\end{equation}
Volumes are measured in string units $\alpha' =
(2\pi)^{-1}$.

One can have stringy deformations of the above field
theory superpotential in the case $ b \,> \, 1$ or $ b \,
\leq \, -1$. The superpotential in this case is

\begin{equation} \label{eq:W_stringy}
W_{\rm stringy}={\Lambda^7\over M_s^6}V_3\, \sum_{b>1
\& b\le -1}
f(b) \, e^{-S_1+(b-1)S_2}.
\end{equation}
where $S_2$ is given by 

\begin{equation} \label{eq:S_2}
{\rm Re}\,(S_2)=3r_3-2r_2
\end{equation}
and $W_{\rm stringy}$ is valid near the quiver locus
$\vert {\rm Re}(S_2)\vert
 \ll 1$. 

Apart from these contributions to the superpotential,
there is the usual term $W_{\rm flux}$ responsible for fixing
complex structure moduli, and
 $W_{\rm gaugino}\sim \Lambda_{SO(8)} e^{-S_3}$
arising from gaugino condensation in pure $SO(8)$ gauge theory on a
divisor $D_6$ at infinity. $D_6$ does not intersect $D_5$, and thus there is no
mixture between quiver fields and instanton effects in $W_{\rm gaugino}$.
Here, $\alpha$ is a number less than one, and $S_3$ is given by

\begin{equation} \label{eq:S_3}
{\rm Re}\, (S_3)=r_2r_3-(1/2) r_3^2 \, .
\end{equation}
The superpotential is a sum of all these effects.
Denoting $x_a = 2{\rm Re}(S_a)$, the regime of validity of this
superpotential is

\begin{equation} \label{eq:valid}
x_{3} \gg x_{1} \gg 1,\quad \vert S_2 \vert \ll 1,
\quad {\rm or
\, equivalently,} \quad r_2\sim (3/2)r_3 \gg r_5 \gg 1
\end{equation}

To simplify the analysis of the vacuum structure of
this system, we set $r_2 =  (3/2)r_3$, and only consider the contribution
from instantons with $b=1$. The superpotential of the system, after
integrating out the fields $Z$ and $X_{ia}$, is

\begin{equation} \label{eq: W_eff}
W_{\rm eff}=W_{\rm flux}+
3\Lambda^{7/3}\kappa^{-2/3}\psi_{3}^{1/3}e^{-S_{1}/3}
+\Lambda_{SO(8)}^3 e^{-\alpha S_3}
\end{equation}
where 
\begin{equation} \label{eq : psi}
\psi_{a} = \kappa^{2}V_{a}, \, \, \, \,
\kappa^2 = M_{pl}^{-2}
\end{equation}

In our regime of validity, we can use the standard
large radius expression for the Kahler potential:

\begin{equation} \label{eq: Kahler}
\kappa^{2} K = -2{\rm log}\Bigl(f_1+
f_2 \sqrt{\psi_a \psi_a^*}\Bigr) \,.
\end{equation}
where $ f_1 $ is the volume of the threefold, and $
f_2 $ is the volume
 of the divisor $ D_5 $, in string units.

Under our approximations, we obtain $f_1$ and $f_2$ in
terms of the fields $x_1$ and $x_3$ as follows:

\begin{equation} \label{eq: f_1}
f_1 = (1/4\sqrt{2})x_3^{1/2}x_1^{1/2}[x_3^{1/2} -
x_1^{1/2} ]
\end{equation}
and
\begin{equation} \label{eq: f_2}
f_2 = (1/2)x_1
\end{equation}

Equipped with $W$ and $K$, we have the supergravity
scalar potential

\begin{equation} \label{eq: V}
V={\rm exp}(\kappa^2 K)
\Biggl(K^{i \bar{j}}W_{eff ;i}W^*_{eff ;
\bar{j}}-3\kappa^2
 W^*_{eff}W_{eff}\Biggr)
+{1\over 2 g_X^2}\sum_{a=1}^3 (D_a)^2
\end{equation}
where the $U(1)$ D-terms are given by:
$$D_1 = -D_2 = -2 \Bigl(\psi^a\partial_{\psi^a}K +
\partial_{x_1}K
 \Bigr),\quad D_3=0.$$

First, we perform an analysis to minimize V with
respect to the fields $\psi_1$ and $\psi_2$. Since these fields do not
appear in the superpotential or its derivatives, their contribution
to the F-term comes from the inverse Kahler metric and derivatives of the
Kahler potential. We study the region of field space where $\alpha_1 =
\psi_{1}\bar \psi_{1}
 \ll \psi_3 \bar \psi_{3} $ and $ \alpha_{2} =
\psi_{2}\bar \psi_{2} \ll
 \psi_3 \bar \psi_{3}$. $V_F$ as a function of
 $\alpha_{1}$ and $\alpha_2$ takes the form:

\begin{equation} \label{eq: Vpsi1}
V_{F}(\alpha_1,\alpha_2) = \kappa^2\Biggl(\frac
{J_1\alpha_1 +
 J_2\alpha_2 + J_3\alpha_1\alpha_2 + J_4} 
{J_5\alpha_1 + J_6\alpha_2 +
 J_7\alpha_1\alpha_2 + J_8}\Biggr) \,  - 3\kappa^2 W
\bar{W}.
\end{equation}

The $J_{i}$ are functions of $\psi_3 \bar \psi_3$ and
$S_1, S_3$. In writing the $J_{i}$, we have used the approximation
$\psi_a \bar \psi_a \sim \psi_3 \bar \psi_3$. We see that 
$J_1,J_2,J_3,J_4$ have mass dimension six, and consist of products of $W$ and its
derivatives. In the limit of $W_{\rm flux} \gg W_{\rm correction}$ where
$W_{\rm correction} =
 3\Lambda^{7/3}\kappa^{-2/3}\psi_{3}^{1/3}e^{-S_{1}/3}
+\Lambda_{SO(8)}^3
 e^{-\alpha S_3} $, we can write

\begin{equation} \label{eq: VWflux}
V_{F}(\alpha_1,\alpha_2) \sim \kappa^2 W_{\rm flux}^2
\Biggl(\frac
 {J_1\alpha_1 + J_2\alpha_2 + J_3\alpha_1\alpha_2 +
J_4} {J_5\alpha_1 + J_6\alpha_2
 + J_7\alpha_1\alpha_2 + J_8} - 3\Biggr) \, .
\end{equation}
where $J_{i}, i = 1 $ to $4$ have been redefined, and
are now dimensionless.  

On the other hand, the D-term contribution is 

\begin{equation} \label{eq:VD}
V_D = \kappa^{-4}g^{-2}[J_9\alpha_1 + J_{10}\alpha_2 +
J_{11}]^2
\end{equation}

For $g^2\kappa^{6}W_{\rm flux}^2 \ll 1$, the D-term
dominates over the F-term, and we can argue that the potential is
minimized at $\alpha_1 = \alpha_2 = 0$.   
Since the F-term contribution is essentially monotonic
as a function of $\alpha_1$ and $\alpha_2$, the minimum will again be
decided by the D-term in the regime
where the F-term and D-terms are comparable. For
 $g^2\kappa^{6}W_{\rm flux}^2 \gg 1$, the F-term
dominates, and the minimum will be decided by
 whether it is monotonically
rising or falling in the regime of validity. Since the
$J_i$ in the numerator and denominator are comparable, this rise
or fall is essentially flat, and we can set $\alpha_1 = \alpha_2 = 0$. 
This also matches with the result in the case of global supersymmetry.

The Kahler metric then simplifies into block diagonal
form, and in particular the inverse entries in $\psi_3, S_1, S_3$
space are unaffected by the $\psi_1$ and $\psi_2$, and a direct analytical
treatment becomes tractable.

We work in the regime where the F-term dominates over
the D-term. Then, the scalar potential becomes (neglecting pure $W_{\rm
correction}$ terms)

$$
V \, \, \sim
e^{\kappa^2K}\Biggl[
\Bigl(-3\kappa^2+K^{i\bar j}\partial_i(\kappa^2 K)
\partial_{\bar
 j}(\kappa^2 K)\Bigr)
\vert W_{\rm flux}\vert^2
$$

\begin{equation} \label{eq : Vapprox}
+ K^{i \bar j} \bigl[ \partial_{i}(\kappa^2 K)W_{\rm
flux}\partial_{\bar
 j}(\bar W_{\rm correction})+cc\bigr]\Biggr]
\end{equation}

Taking the open string field
$V_3$ to be stabilized below $M_{{\rm Planck}}$ we get $\vert \psi_3
\vert \ll 1$. Also, the regime of validity
of the model is $x_3 \gg x_1$.

Evaluating the inverse Kahler metric and keeping to
lowest powers of
 $\vert \psi_3 \vert$ and $x_1 / x_3$, one obtains

$$
e^{\kappa^2K}\kappa^4(\partial_{S_1}K)^2 K^{S_1
\bar{S_1}} W_{\rm flux}^2
 \, \, \sim \, \,  e^{\kappa^2K}\kappa^2 (.33)W_{\rm
flux}^2 \, ,
$$

$$
e^{\kappa^2K}\kappa^4(\partial_{S_1}K)(\partial_{S_3}K)
K^{S_1
 \bar{S_3}} W_{\rm flux}^2 + c.c. \, \,  \sim \, \, 
e^{\kappa^2K}\kappa^2
 (.5)W_{\rm flux}^2 \, ,
$$

$$
e^{\kappa^2K}\kappa^4(\partial_{S_3}K)^2 K^{S_3
\bar{S_3}} W_{\rm flux}^2
 \, \, \sim \, \,  e^{\kappa^2K}\kappa^2 (.25)W_{\rm
flux}^2 \, ,
$$
while other contributions to $e^{\kappa^2K} K^{i\bar
j}\partial_i(\kappa^2 K) \partial_{\bar
 j}(\kappa^2 K)W_{\rm flux}^2$ contain positive powers
of $\vert
 \psi_3 \vert$ and $x_1/x_3$ and are thus further suppressed.

One thus obtains 

\begin{equation} \label{eq:V1}
e^{\kappa ^2K}
\Bigl( -3\kappa^2 + K^{i\bar j}\partial_i(\kappa^2 K)
\partial_{\bar
 j}(\kappa^2 K)\Bigr)  \, \, < \, \, 0 \, \, .
\end{equation}

On the other hand, $ K^{i \bar j} \bigl[
\partial_{i}(\kappa^2
 K)W_{\rm flux}\partial_{\bar j}(\bar W_{\rm
correction})+cc\bigr] $  gives

$$
e^{\kappa^2 K} \Bigl[ \Lambda^{7/3} \kappa^{4/3}\vert
\psi_3 \vert
 ^{1/3} e^{-x_1/6}(1 + 2(x_1/x_3) + ...) \cos
(\theta/3 - \text{Im}S_1 /3)
$$
\begin{equation} \label{eq:second}
+ \kappa^2 \Lambda_{SO(8)}^{3} \alpha \, e^{-\alpha
x_3}x_3 \cos(\alpha
 \text{Im} S_3) \Bigr]W_{{\rm flux}}
\end{equation}
where $\theta$ is the phase of $\psi_3$. Setting 
 $\theta/3 - \text{Im}S_1/3 = \pi$ and $\alpha
\text{Im} S_3 = \pi$, we get
 a negative contribution from this term also.

One thus obtains a negative scalar potential in the
regime of calculability of the theory. As $x_3$ and $x_1$ grow
large, $e^{\kappa^2K} \sim
 x_3^{-2}x_1^{-1}$ damps out the scalar potential, and $V$ goes to zero.
Since the potential is also bounded below as long as the model
is well-defined, one obtains an $AdS$ minimum. We note that the metastable minimum 
of the system may lie outside the regime of calculability.

%%%%%%%%%%%%%%%%%%%%%%%%%%%%%%%%%%%%%%%%%%%%%%%%%%%%%%%%%%%%%%%%%%%%%
%%%%%%%%%%%%%%%%%%%%%%%%%%%%%%%%%%%%%%%%%%%%%%%%%%%%%%%%%%%%%%%%%%%%%

%%%%%%%%%%

\end{document}